\documentstyle[prl,aps,epsf]{revtex}
\begin{document}
\title{Temporal intermittency and cascades in shell models of turbulence}
\author{P. D. Ditlevsen$^{\dag}$\\
National Center for Atmospheric Research,\\ Table Mesa Dr., Boulder, CO-80307.}
\date{December 8, 1995}
\maketitle
\begin{abstract}
The 2d and 3d like Gletzer, Okhitani and Yamada (GOY) shell models
are examined. The 2d like model shows a transition from statistical
quasi-equilibrium to cascade of enstrophy as a function of the
spectral ratio of energy to enstrophy. The transition is related to
the ratio of time scales, corresponding to eddy turnover times,
between shells. The anomalous scaling, giving rise to non-linear
scaling functions, is also connected to the ratio of eddy turnover
times. This is illustrated in a simple stochastic model, where
the structure function, $\zeta (q)$, becomes independent of $q$.
In the 3d like model the multiscaling is also influenced by the
existence of a second non-positive definite inviscid invariant,
the helicity. 
\end{abstract}

The problem of understanding the scaling properties of velocity correlations 
in isotropic and homogeneous turbulence is still largely 
unsolved. 
The only exact
derived property is the Kolmogorov scaling law for the third order
moments, $C_3(r)=-4\epsilon r/5 +6 \nu dC_2(r)/dr$, where 
$C_q(r)\equiv \langle \Delta v(r)^q\rangle = \langle |v(x+r)-v(r)|^q\rangle,
r$ is the distance in the fluid and $\langle .\rangle$ is the ensamble average.
This relation is not closed, since it contains both second and third order
correlation functions. In the inertial range defined by, $L \gg r \gg \lambda$,
where $L$ is the outer scale and $\lambda$ the inner or Kolmogorov dissipation scale, the second term on the 
right hand side is 
negligible and the scaling relation, $C_3(r)\sim r^{\zeta (3)}$, with $\zeta (3)=1$,
holds. The classical Kolmogorov theory (K41) based on dimensional arguments
states that $C_q(r)\sim r^{\zeta(q)}$, with $\zeta(q)=q/3$. 
It was noted by Landau shortly after the theory was presented that the
energy dissipation could vary so much as to alter the scaling laws. This
was incorporated in a refined version of Kolmogorovs theory (K62),
in which a dependence on $L/r$, the ratio of the
outer scale to the scale in the inertial range was incorporated.

It has been
seen in numerous experiments that $\zeta(q)$ is a weakly non-linear
function of $q$, different from the 
Kolmogorov predictions for $q$ different from 3. It follows from the 
H{\o}lder inequality that
$\zeta (q)$ is a convex of $q$, thus
$\zeta (q) > q/3$ for $q< 3$ and $\zeta (q) < q/3$ for $q> 3$. 
The deviation of the scaling from the K41 prediction is referred to as
intermittency corrections. It is widely attributed to the fact that the energy
dissipation in fully developed turbulence
is indeed highly inhomogeneous, basically taking place on lower
dimensional subsets, 
filaments, 
of the flow field.

Numerical studies of the GOY shell model \cite{GOY} of turbulence have been popular resently,
mainly because this model, in which scaling relations completely equivalent
to those of turbulence, also shows intermittency corrections to the K41
- equivalent - predictions. The GOY model structually resembles the
spectral form of the Navier-Stokes equation, but there is no spatial
fields associated with the wave components in the GOY model. 
The intermittency corrections in the scaling properties are thus 
associated with temporal intermittency which can only be weakly linked to
spatial intermittency through the Taylor hypothesis. 
The connection between the spatial intermittency of turbulence and 
temporal intermittency of the GOY model is not clear, but the hope
is that understanding the latter can shed light upon the former. This
note is about the latter.

In the GOY model the spectral domain is represented as shells,
each of which is defined by a wavenumber $k_n = \lambda^n$, where
$\lambda$ is a scaling parameter defining the shell spacing.
There are $2 N$
degrees of freedom, where $N$ is the number of shells, namely the
generalized complex shell velocities, $u_n$ for $n=1,N$.
The dynamical equation for the shell velocities is,

\begin{equation}
\dot{u}_n=i k_n (u_{n+2}u_{n+1}-\frac{\epsilon}{\lambda}u_{n+1}u_{n-1}
+\frac{(\epsilon -1)}{\lambda^2}u_{n-1}u_{n-2})^* -\nu k_n^2 u_n + f \delta_{n,n_0},
\label{dyn}
\end{equation}

where the first term represents the non-linear wave interaction or
advection, the second term is the dissipation, and the third term the
forcing, where $n_0$ is some small wavenumber. The boundary conditions
are, $u_{-1}=u_0=u_{N+1}=u_{N+2}=0$. The model has two quadratic
inviscid invariants, which are constants of motion in the case $f=\nu=0$, $E=
\sum_n |u_n|^2$, referred to as the energy, and $H=\sum_n (\epsilon -1)^{-n}
|u_n|^2$.
For $\epsilon<1$ $H$ is non-positive
definite, referred to as the helicity, and the model is thought of as
modeling 3d turbulence. For $\epsilon>1$ $H$ is positive definite,
referred to as the enstrophy, and the model is meant to resemble 2d turbulence.
Dimensional arguments similar to those of K41 can be applied to the GOY 
model assuming dissipation of one of the conserved quantities. For the
2d type model, $1<\epsilon<2$, enstrophy will be dissipated, 
and an additional large
scale drag term, $-\nu' k_n^{-2}u_n$, is added to (1) 
in order to remove the energy. The reason
for the enstrophy and not the energy being dissipated is the usual that
the dissipation of energy is $k_n^{-\alpha}$ times the dissipation 
of enstrophy at shell $n$, thus negligible for $\alpha > 0$ and $n\rightarrow
\infty$. The Kolmogorov scaling for the 2d type model then becomes,
$\langle |u_n|\rangle \sim k_n^{-(\alpha+1)/3}$. 
Note that this is an unstable fixed
point of (1) for $f=\nu=0$, describing a cascade. As pointed out
by Aurell et al. \cite{Aurell} in the case of $\alpha=2$ this scaling
coincides with the scaling that would be obtained in a statistical
equilibrium, in which the enstrophy is distributed evenly among the
degrees of freedom of the system, $\langle |u_n|\rangle \sim k_n^{-\alpha /2}$. The
$\alpha =2$, corresponding to $\epsilon =5/4$, case is a borderline 
case between models showing cascade, for $\alpha < 2$, and models
showing statistical (quasi-)equilibrium, for $\alpha > 2$. 
In order to show this a series of numerical model runs for 
different values of $\epsilon$ have been performed, details of
which are reported elsewhere \cite{ditlev}. 

\begin{figure}[htb]
\epsfxsize=8cm
\epsffile{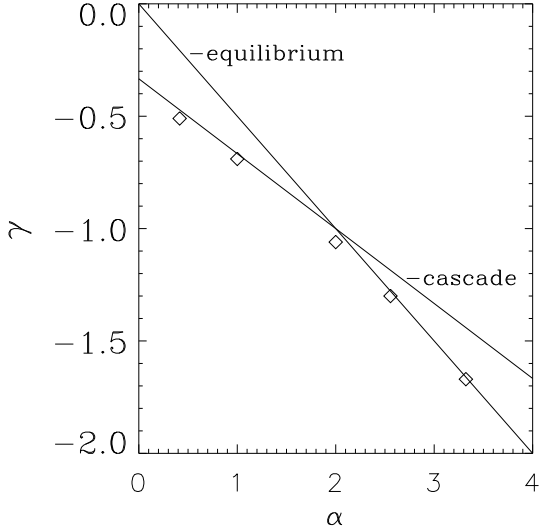}
\caption[]{ 
Results of runs of the 2d like GOY model for different
values of $\epsilon$. Shown is
the spectral slope $\gamma=-\log (\langle u\rangle)/
\log (k)$ as function of $\alpha = -\log (\epsilon -1)/\log (\lambda)$.
The lines are the Kolmogorov scaling for enstrophy
cascade and equipartitioning of enstrophy, respectively.
For the 2d like GOY models an additional dissipation 
of the form, $-\nu' k_n^{-2} u_n$, is applied in order to
remove energy at small wave number shells.
The model is run with $n=50, k_0= \lambda^{-4}, \lambda=2,
f_n= 5 \times 10^{-4} \times (1+i) \delta_{n,15},
(\epsilon =11/10,\nu = 5 \times 10^{-27}, \nu'=100),
(\epsilon =117/100,\nu = 5 \times 10^{-27}, \nu'=100),
(\epsilon =5/4,\nu = 5 \times 10^{-25}, \nu'=100),
(\epsilon =3/2,\nu = 5 \times 10^{-23}, \nu'=100),
(\epsilon =7/4,\nu = 5 \times 10^{-23}, \nu'=100),
(\epsilon =2,\nu = 5 \times 10^{-20}, \nu'=100).$
The models show a cross-over at $\alpha=2$ between
statistical equilibrium, $\alpha>2$, and enstrophy cascading,
$\alpha<2$.
}
\end{figure}

Figure 1 shows the
spectral slopes obtained. The two lines shows the slopes corresponding
to cascade and statistical equilibrium respectively, the diamonds
are the numerical results. The figure clearly shows a cross-over
from one type of behavior to the other. The reason for this
transition is related to the scaling of typical timescales, or
eddy turnover times, for the different shells. The eddy turnover 
time at shell $n$ is given as $\tau_n = (k_n \langle |u_n|\rangle)^{-1}
$ as seen from (1) or dimensional arguments. The eddy turnover time
then becomes, $\tau_{\mbox{\small{cascade}}}\sim k^{(\alpha -2)/3}$ and 
$\tau_{\mbox{\small{equilibrium}}}\sim k^{(\alpha -2)/2}$ respectively. Thus
for $\alpha >2$ the eddy turnover times in both cases increase with
wave number, and equilibration via inverse enstrophy transfer, from
large wave numbered shells to small wave numbered shells takes place.
For $\alpha < 2$ the situation is reversed, and a statistical 
equilibrium can newer be achieved. Similar results have been found
by Yamada and Ohkitani \cite{OY1} for a slightly different set of
GOY models, with only one positive definite inviscid invariant, 
reffered here to as 3D type models. 

In the 3d type models the situation is totally different, in this case
the second inviscid invariant, the helicity, is not cascaded even 
though the ratio of absolute value of the helicity to the energy
at shell $n$ grows exponentially with $n$ as is the case for the
enstrophy. The reason is that the helicity 
has alternating signs for even and odd numbered shells. Therefore
there will be a net production of (positive sign) helicity at
odd numbered shells due to the dissipation, and no net flow of
helicity from small wave number shells to large wave number shells
is necessary. It has been demonstrated numerically  that helicity
is indeed the quantity to be cascaded in the (pathological) case
of hyperviscosity only active on the outermost shell \cite{ditlev2}.
In the usual 3d case energy will be cascaded, with the
resulting spectral scaling, $\langle |u|^3\rangle \sim k^{-1}$.  

Both the 2d and the 3d models shows intermittency corrections
to the Kolmogorov scaling depending on $\epsilon$ and $\lambda$.
In this note I will suggest two different mechanisms in play
in the 3d case, where only the one is in action in the 2d case. 

In the 3d case it is observed
that the structure function depends only on $\epsilon$ and
$\lambda$ in the combination, $-\log (1-\epsilon)/\log (\lambda)=
\alpha$ \cite{Kadanoff}. With $\delta\zeta (q)\equiv \zeta (q)-
q/3$, $\delta\zeta (q)$ increases in absolute value when 
$\alpha$ increases. I suggest that this is due to differences in
the ratio of helicity production and helicity elimination
by dissipation at neighboring shells in the beginning of the 
dissipative subrange where scaling still approximately holds or
where extended self-similarity can be applied \cite{benzi}.
This ratio, $r$, can be estimated as 
$r\equiv (\Delta H_{n-1}+\Delta H_{n+1})/2\Delta H_n \sim
(\lambda^{-\alpha +2/3}+\lambda^{\alpha -2/3})/2$, 
which for $(\alpha , \lambda)=(0.5,2),(1,2),(2,2)$ gives
$1.007, 1.03, 1.46$ respectively. So there is the biggest
"mis-match", or non-cancellation,  in the case $(2,2)$ where the biggest non-linearity
in the structure function is observed. That these two things
should be related is consistent with the findings of a
moderated GOY model (model 3) by Benzi et al. \cite{Bob}, where 2
copies of the GOY model are coupled. In this model the helicity 
takes the form, $H=\sum_n k_n^{\alpha}(|u_n^+|^2-|u_n^-|^2)$, 
where $u_n^+$ and $u_n^-$ are the two complex variables in shell $n$.
In this model the helicity production and elimination will
- on average - exactly balance, thus no dependence of the 
structure function on $\alpha$ should be expected in agreement
with the numerical findings. 

In the standard 3d case, $\alpha =1$, the structure function is 
still non-linear even though the above defined $r$ is close to
1. This is suggested to be due to the difference in time scales
between the large and the small scales.
This effect is expected also to determine the anomalous scaling
in the 2d case, where only the
dissipation of enstrophy plays a role. In the 2d case numerical
studies suggests that the absolute size of $\delta \zeta (q)$
increases as $\alpha$ decreases from 2 to 0 \cite{ditlev}. 
This suggests
that intermittency is enhanced with growing difference between
eddy turnover times from one shell to the next, which goes
as $\tau_{n+1}/\tau_n=\lambda^{(2-\alpha)/3}$. 

\begin{figure}[htb]
\epsfxsize=8cm
\epsffile{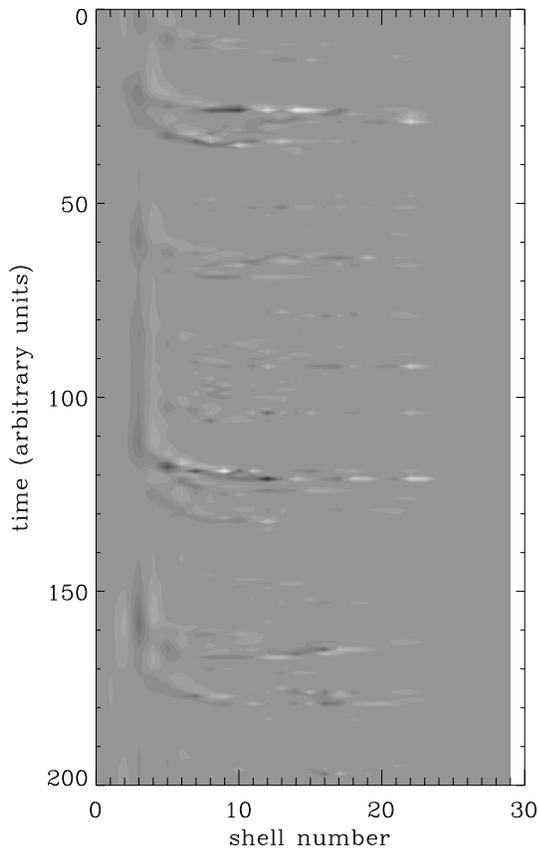}
\caption[]{ 
The non-linear transfer of energy from one shell to the next in the
3d like GOY model with $\epsilon = 1/2$, all parameter values are
standard, forcing is on shell number 4. The line segments pointing
from shell $n$ at time $t$ to shell $n+1$ at time $t+\Delta t$
symbolize forward transfer, with magnitude proportional to the
thickness of the line segment. Line segments pointing from
shell $n+1$ at time $t$ to shell $n$ at time $t+\Delta t$ symbolizes
backward transfer.
}
\end{figure}

Figure 2 shows the non-linear transfer, $\Delta_n = 
k_{n-1}Im(u_{n-1}u_nu_{n+1})$ of energy from shell $n-1$ to shells $n$
and $n+1$ in the 3D case. 
The abscissa is the shell number and the ordinate is time,
read from top to bottom, the figure is composed of line-segments
connecting shells $n$ at time $t_i$ and shells $n+1$ at time
$t_i+\Delta t$, symbolizing an energy transfer from shell $n$ 
in the time-interval $[t_i;t_i+\Delta t]$. The thickness of the
line-segment is proportional to the size of the transfer. Line-segments
going from $n+1$ at time $t_i$ to $n$ at time
$t_i+\Delta t$ symbolizes the inverse energy transfer. The 
figure shows that the transfer is temporally intermittent and
occurring in bursts \cite{mhj}, with
the transfer being faster as the bursts propagate to higher
wave-number shells. From this the interpretation is straight forward;
The residence time for a burst at a given shell is proportional
to the eddy turnover time resulting in a more and more intermittent
transfer as the eddy turnover time decreases. In order to illustrate 
how this kind of behavior can lead to anomalous scaling behavior,
consider the following linear stochastic model, which is an extreme case.

Let $x_n$ be a stochastic variable representing 
the energy at shell $n$, governed by the 
dynamical equation;

\begin{equation}
x_n(t+1)=  p_{n-1}(t)x_{n-1}(t)-p_n(t) x_n(t) -\nu k_n^2 x_n(t) +f\delta_{n,1},
\end{equation}

where $p_n$ is a stochastic variable which is 
1 with probability $\tau_N/\tau_n$ and 0 
with probability $1-\tau_N/\tau_n$, thus it represents a transition
probability of energy transfer from shell $n$ to shell $n+1$. 
The boundary conditions are $x_{-1}=p_N=0$. The
average residence time is then simply proportional to the 
eddy turnover time. It is easily seen that $x_n$ is always positive.

\begin{figure}[htb]
\epsfxsize=8cm
\epsffile{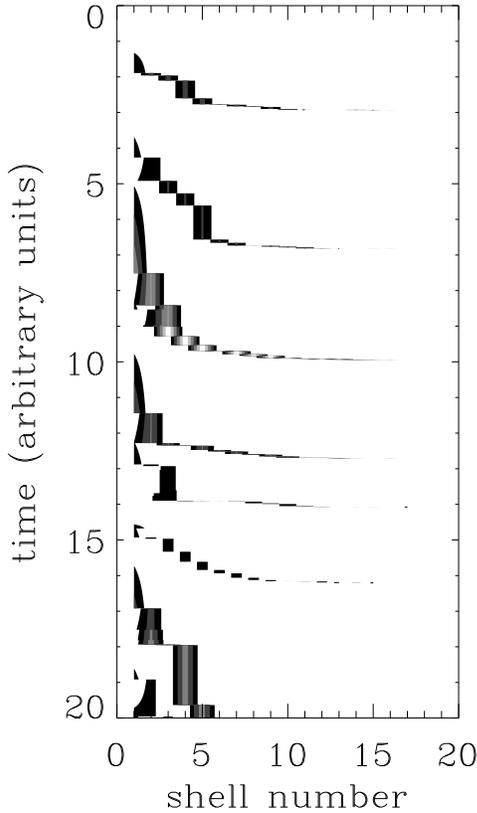}
\caption[]{ 
The result of a simulation of the simple stochastic model, $x_n$
is plotted as a function of time. The intermittent transfer 
is similar to what is seen in the GOY model, figure 2.
}
\end{figure}

Figure 3 is similar to figure 2, but shows
the values of $x_n$. There
is obviously no inverse transfer in this model. By comparing figures
2 and 3 one sees that the intermittent structure of the 
transfers are similar. The energy will in this model scale
inversely proportional to the eddy turnover time, 
so that if we identify $x_n$ with the energy/enstrophy of
the GOY model $E_n=k_n^{\alpha}|u_n|^2$, the scaling of
$\langle |u_n|\rangle \sim k_n^{-(\alpha +1)/3}$ is the same as for the GOY model.
However, the structure function changes completely. It is
readily calculated, with the result; $\langle x_n^q\rangle
\sim k_n^{(\alpha -2)/3} x^q$, where $\tau_n = k_n^{(\alpha -2)/3}$ is
the eddy turnover time at shell $n$ and $x = f \tau_1$ is the 
average energy input into the system during one large eddy turnover time.
This means that the structure function, $\zeta (q)$, becomes 
independent of $q$. $q$ will only show up in the offset, $q \log (x)$ in
a $\log \langle x_n^q\rangle$, $\log (k_n)$ plot. This is
illustrated in figure 4, showing the result of a numerical simulation.

\begin{figure}[htb]
\epsfxsize=8cm
\epsffile{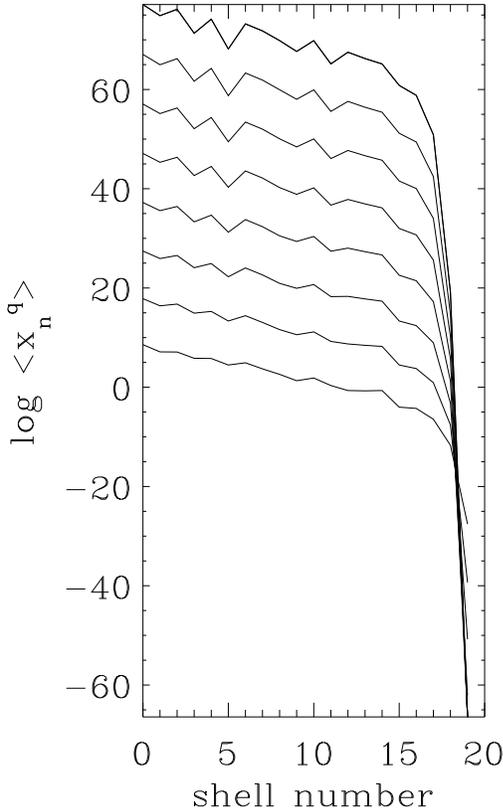}
\caption[]{ 
Anomalous scaling behavior of the simple stochastic model. 
$\log (\langle x^q\rangle)$ vs. $\log (k)$ for $q=1,8$.
The
spectral slope 
does not depend on $q$, So that $\zeta (q) = (\alpha -2)/3$.
The model was iterated for $10^4$ large eddy turnover times
with $N=20, f=0.05, \nu =10^{-11}$.
}
\end{figure}

Obviously this linear stochastic model cannot reproduce the scaling
exponents of the GOY model or of real turbulence but it qualitatively
illustrates the effect of temporal intermittency.

In conclusion, 
the behavior of the 2d like GOY model shows
either statistical equilibrium or cascade of enstrophy depending on the
ratio of the eddy turnover timescales between the shells. 
The
dynamics and the multiscaling of the 3d like GOY model depends 
on the existence of
the second non-positive definite inviscid invariant, helicity,
through the
way the helicity is dissipated in the model.
In both cases, the multiscaling is an effect of temporal intermittency
also originating from differences in eddy turnover times at
the different scales, thus the 2d model with $\alpha =2$ shows
no anomalous scaling. The effect is illustrated in a simple 
stochastic model.
The scaling behavior of this simple model does not correspond to
what is seen in the GOY model or in real turbulence. However, 
the model points to a mechanism of how the temporal intermittency can lead to
anomalous scaling behavior. In order to further clarify the
relationship between the spatial intermittency observed in 
turbulence and the temporal intermittency seen in these simple
models, it would be interesting to see if the structures seen
in the time - wave-number domain, figure 1, can be identified
in direct numerical simulations of the Navier-Stokes equation.

Acknowledgements: 
It's a pleasure to thank R. M. Kerr and J. Herring 
for valuable discussions and NCAR for 
hospitality and financial support. The work was granted
by the Carlsberg foundation.

$^{\dag}$ On leave from The Niels Bohr Institute, Department for Geophysics, 
University of Copenhagen, Juliane Mariesvej 30, DK-2100 Copenhagen O, Denmark.
%\begin{references}


\begin{thebibliography}{99}
\bibitem{GOY} M. Yamada and K. Okhitani, J. Phys. Soc. of Japan 56,
4210 (1987); Progr. Theo. Phys. 79, 1265 (1988); Phys. Rev. Lett..
60,983 (1988)
\bibitem{Aurell}E. Aurell, G. Boffetta, A. Crisanti, P. Frick, G. Paladin and A. Vulpiani, Phys. Rev. E, 50, 4705, (1994).
\bibitem{ditlev}P. D. Ditlevsen and I. A. Mogensen, Phys. Rev. E, 53, 4785, 1996.
\bibitem{OY1} M. Yamada and K. Okhitani, Phys. Lett. A, 134, 165, 1988.
\bibitem{ditlev2}P. D. Ditlevsen, Phys. Fluids, 9, 1482-1484, 1997.
\bibitem{Kadanoff}R. Benzi, L. Kadanoff, D. Lohse, M. Mungan and J. Wang, Phys. Fluids 7, 617, 1995.
\bibitem{benzi}R. Benzi, S. Ciliberto, C. Baudet, G. R. Chavarria, Physica D. 80, 385, 1995.
\bibitem{Bob}R. Benzi, L. Biferale, R. M. Kerr and E. Trovatore, preprint chao-dyn/9510010 in PBB for non-linear science.
\bibitem{mhj}M. H. Jensen, G. Paladin and A. Vulpiani, Phys. Rev. A, 43, 798, 1991. 
%\bibitem{b+k}L. Biferale and R. Kerr, preprint chao-dyn/9508007 in PBB for non-linear science.
%\end{references}
\end{thebibliography}
\end{document}